\documentclass[a4paper,UKenglish,cleveref, autoref, thm-restate]{article}
\usepackage{graphicx}
\usepackage{xcolor}      
\usepackage{pifont}      
\usepackage{subcaption}  
\usepackage{amsmath}     
\usepackage{amsthm}      
\usepackage{amssymb}     
\usepackage{enumitem}    
\usepackage{listings}    
\usepackage{hyperref} 
\usepackage{cleveref}

\usepackage{amsthm}

\theoremstyle{plain}
\newtheorem{theorem}{Theorem}[section]
\newtheorem{lemma}[theorem]{Lemma}
\newtheorem{corollary}[theorem]{Corollary}

\theoremstyle{definition}
\newtheorem{definition}[theorem]{Definition}
\newtheorem{example}[theorem]{Example}

\theoremstyle{remark}

\title{Color Structures and the Monotone Satisfiability Problem with Bounded Variable Occurrence} 

\author{Ronald {de Haan} \and Hannah Van Santvliet}

\newcommand{\n}[1]{\ensuremath{\overline{#1}}}
\usepackage{tikz}
\usetikzlibrary{automata, positioning, arrows.meta, decorations.pathmorphing}
\usepackage{comment}

\begin{document}
\maketitle
\begin{abstract}
We study \textsc{Monotone 3-Sat-$(\leq k,1)$}, a restricted variant of the \textsc{Satisfiability} problem where clauses consist of three variables and are monotone (every clause contains either only unnegated or only negated variables) with up to $k$ positive and exactly one negative occurrence per variable in the formula.
We resolve a challenge posed by Darmann and Döcker (On simplified NP-complete variants of \textsc{Monotone} 3-\textsc{Sat}, Discrete Applied Mathematics 292:45--58, 2021) by proving that for~$k\in \{3,4\}$, the problem is trivial in the sense that every instance satisfying the given restrictions is satisfiable. This result closes the remaining gap in a dichotomy theorem: Triviality for $k\in \{1,2\}$ follows by a result by Tovey (A simplified NP-complete satisfiability problem, Discrete Applied Mathematics 8(1):85--89, 1984), while NP-completeness for~$k\geq 5$ was shown by Darmann and Döcker. To obtain our result, we introduce the notion of \emph{color structures}  and show that a satisfying assignment can always be constructed in $\mathcal{O}(n \cdot m)$ time, where $n$ and $m$ denote the  number of negative and positive clauses of the input formula, respectively. 
\end{abstract}

\section{Introduction}

The Boolean satisfiability problem (\textsc{Sat}) is a central decision problem in theoretical computer science. Given a Boolean formula in conjunctive normal form (CNF), the problem asks whether there exists an assignment that satisfies all clauses. Cook’s seminal result established that \textsc{Sat} is NP-complete \cite{cook}, and the problem has since served as a benchmark for studying computational hardness.

A common approach to gaining a finer understanding of the complexity of \textsc{Sat} is to study restricted subclasses of CNF formulas. By imposing syntactic or structural constraints,  the resulting satisfiability problem may become tractable. 

A well-known example for such a jump is the restriction to clauses of size two: While 3-\textsc{Sat} is NP-complete, 2-\textsc{Sat} is solvable in polynomial time \cite{AspvallPlassTarjan1979}.
Further restrictions, such as bounding the number of variable occurrences, can render the problem trivial in the sense that every restricted formula is either always satisfiable or always unsatisfiable.
Tovey showed that any CNF formula with clause size $r$ in which each variable occurs at most $r$ times is always satisfiable \cite{tovey}.
More recent work in this direction---aimed at identifying tractable or in particular trivial subclasses---has explored restrictions on the graph representation of formulas \cite{pilz} and on the number of potential conflicts arising from truth value assignments \cite{scheder}.

Among the restricted variants of \textsc{Sat}, monotone formulas---formulas in which each clause contains either only positive or only negative literals---have received considerable attention. Since \textsc{Monotone 3-Sat} is NP-complete \cite{gold, li},  monotonicity alone does not suffice to render satisfiability easy. Consequently, additional restrictions are required to identify subclasses where satisfiability is guaranteed.

In the monotone setting, particular attention has also been paid to formulas where the number of occurrences of each variable is explicitly bounded. Darmann, D\"ocker, and Dorn~\cite{DDD} show that \textsc{Monotone 3-Sat} remains hard if each variable appears exactly four times. 
Darmann and D\"ocker \cite{janoschPaper} distinguish positive and negative occurrences of the variables. They study the class \textsc{Monotone 3-Sat}-$(p,q)$, consisting of monotone 3-CNF formulas in which each variable occurs exactly~$p$ times positively and~$q$ times negatively, and obtain hardness results for  every fixed pair $(p, q)$ with $p \geq 2$ and $q \geq 2$, as well as for the case that each variable appears exactly three times positively and once negatively, or three times negatively and once positively. 

They also show that \textsc{Monotone 3-Sat}-$(k,1)$ is NP-complete for $k \geq 5$, and pose the question of the complexity of \textsc{Monotone 3-Sat}-$(k,1)$ with $k < 5$. By a result of Tovey~\cite[Theorem 2.4]{tovey}, the problem is trivial for $k\leq 2$, but the cases $k\in \{3,4\}$ remained open. This question is particularly intriguing as it probes the precise threshold between hardness and guaranteed satisfiability within a structurally simple class of formulas. 

In this paper, we resolve the open question of Darmann and D\"ocker by proving that all instances of \textsc{Monotone 3-Sat}-$(k,1)$ with $k \leq 4$ are satisfiable, establishing that the satisfiability problem for this class is trivial. Our proof introduces a new combinatorial framework, called a \emph{color structure}, which captures the interaction between positive and negative clauses under bounded variable occurrences. We present an algorithm that iteratively builds such a color structure from a given monotone formula with bounded occurrences running in $\mathcal{O}(n \cdot m)$ time, where $n$ and $m$ denote the number of negative and positive clauses of the input formula, respectively. We prove correctness of the algorithm and show that a satisfying assignment can be extracted directly from the resulting color structure.

\section{Definitions and Notation}

Let $x_1, x_2 , \dots$ be Boolean variables taking values in $\{{\texttt{true},\texttt{false}}\}$. A \emph{literal} is either a variable~$x_i$ or its negation~$\n{x_i}$. A \emph{clause} is a disjunction of literals, and a formula is in \emph{conjunctive normal form} (CNF) if it is a conjunction of clauses. If every clause in a CNF formula contains exactly $k$ literals, the formula is called a $k$-\textsc{Sat} formula. A \emph{variable occurrence} is a specific appearance of a variable in a clause of the formula, possibly negated, and possibly appearing multiple times. A CNF formula is \emph{monotone} if every clause consists entirely of positive literals or entirely of negative literals. Accordingly, the clauses of a monotone CNF-formula partition into \emph{positive clauses}, containing only unnegated variables, and \emph{negative clauses}, containing only negated variables.

We further bound the number of variable occurrences. The class \textsc{Monotone 3-Sat}-$(p,q)$ consists of all monotone 3-\textsc{Sat} formulas in which each variable occurs exactly~$p$ times positively and exactly~$q$ times negatively. Analogously, \textsc{Monotone 3-Sat}-$(\leq p,\leq q)$ consists of all monotone 3-\textsc{Sat} formulas in which each variable occurs at most~$p$  times positively and at most~$q$ times negatively. Throughout this paper, we consider instances of \textsc{Monotone 3-Sat}-$(\leq k,1)$ or \textsc{Monotone 3-Sat}-$(k,1)$ for $k \in \{3,4\}$.

To simplify such formulas, we encode the information carried by negative clauses by a \emph{color} (denoted by a letter) and say that variables in the same negative clause \emph{have the same color}. If two (unnegated) variables have the same color, we call them \emph{relatives}. A variable is not its own relative. Since the negated literals occur exactly once, we omit them and carry this information 
into the positive clauses via their respective colors. 
The simplified formula consisting of positive (and colored) clauses only is called a \emph{colored} formula. Note that coloring a formula imposes an order on the variables that appear in the formula. We say that a variable is \emph{smaller} than another variable if the letter corresponding to its color appears earlier in the alphabet or, if both variables have the same color, if the index of the variable is smaller. If we want to explicitly highlight that the order and letter is not relevant, we use $u_i, v_j$ and $l_s$ as variable names.

\begin{example}
For instance, in the formula $$(\n{x_1} \lor \n{x_2} \lor \n{x_{3}}) \land (\n{x_4} \lor \n{x_5} \lor \n{x_{6}}) \land (\n{x_7} \lor \n{x_8} \lor \n{x_{9}}) \land (x_1 \lor x_3 \lor x_{8}) \land (x_1 \lor x_4 \lor x_{7}) \land (x_7 \lor x_8 \lor x_{9})$$ the variables $x_1, x_2, x_3$ are assigned to the color $a$, the variables $x_4, x_5, x_6$ have color $b$ and the variables $x_7,x_8,x_9$ have color $c$. The colored formula consists of the positive (and now colored) clauses $(a_1 \lor a_3 \lor c_2),
(a_1 \lor b_1 \lor c_1)$ and $
(c_1 \lor c_2 \lor c_3)$. The variable $a_3$ has the two relatives $a_1$ and $a_2$. The relative $a_1$ occurs twice in the colored formula, whereas the variable $a_2$ does not occur. 
\end{example} 
Central to our approach is the notion of \emph{guards}. A \emph{guard} is a variable that is explicitly chosen to be assigned to the value \texttt{false}. Intuitively, the set of guards represents the choices we make in constructing the assignment. A negative clause is satisfied if and only if at least one of its variables is assigned the value \texttt{false}. We may therefore assume, without loss of generality, that any satisfying assignment sets exactly one variable per negative clause to false: if more than one variable in some negative clause was set to \texttt{false}, the assignment of the excess false variables could be flipped to \texttt{true} without harming satisfiability. Therefore, the negative clauses of a formula are satisfied if there is exactly one guard for each color. We say that a variable is \emph{unguarded} if it is not a guard and there is no relative of the variable that is a guard. To reason about a formula and a variable assignment simultaneously, we introduce a combined representation called a \emph{color structure}. A colored monotone 3-\textsc{Sat}-$(\leq k, 1)$ formula and a corresponding assignment can be translated into a color structure and the other way around.

\begin{definition} A \emph{color structure} of a colored monotone 3-\textsc{Sat} formula is a graph $(G \cup N,E )$ with two types of variable vertices $G$ (guards) and $N$ (non-guards), together with a set of labeled edges $E$. An element $\{u_i,v_j\}^{l_s}$ in $E$ consists of variable vertices $u_i$ and $v_i$ that form an edge and a variable $l_s$ as edge label. A color structure must satisfy the following properties:
\begin{enumerate}
    \item If two variables of a clause are contained in $G \cup E$, then the remaining third variable of the clause is an edge label and forms a labeled edge with the remaining two variables of the clause.
    \item No two guards have the same color.
    \item Variables that are contained in $G$ are not contained in $N$ and are no edge labels of edges in $E$.
\end{enumerate}
Given a color structure $(G \cup N,E )$, the \emph{induced assignment} sets all variables appearing in $G$ to \texttt{false} and  all remaining variables to \texttt{true}. 
\end{definition}
Note that not all variables of the formula need to occur in $G \cup N$ and that for a clause containing two variables that are both not guards, we may  freely choose which one to place as an edge label and which as a non-guard node, since the disjunction $\lor$ of the formula is commutative. 
\begin{example}\label{example:colorsmall}
Given the (colored) formula $(a_1 \lor a_3 \lor c_2) \land (a_1 \lor b_1 \lor c_1) \land
(c_1 \lor c_2 \lor c_3) \land (b_1 \lor c_1 \lor c_3)$, we construct the color structure $(G=\{a_1, c_1\} \cup N=\{c_2,c_3\},E=\{\{a_1,c_2\}^{a_3},
\{a_1,c_1\}^{b_1},
\{c_1,c_3\}^{c_2},
\{c_1,c_3\}^{b_1} \})$  (see Figure \ref{figure:colorexamplesmall}). Since there is no guard that of color $b$, the variable $b_1$ is unguarded. The induced assignment of the color structure sets~$a_1$ and~$c_1$, corresponding to the guard variables $x_1$ and $x_7$, to \texttt{false} and all remaining variables to \texttt{true}.
\begin{figure}
\centering
    \begin{tikzpicture}[shorten >=0pt,node distance=2cm,on grid,auto] 
   \node[state] (q_0) {$a_1$};
   \node[state] (q_1) [right=of q_0] {$c_1$};
   \node (q_2) [left=of q_0] {$c_2$};
   \node (q_3) [right=of q_1] {$c_3$};
   \node (q_4) [below right=of q_1] {$c_3$};
\path
(q_0) edge node {$a_3$} (q_2)
(q_1) edge node {$c_2$} (q_3)
(q_0) edge node {$b_1$} (q_1)
(q_1) edge node {$b_1$} (q_4);
\end{tikzpicture}
\caption{A color structure of the formula $(a_1 \lor a_3 \lor c_2) \land (a_1 \lor b_1 \lor c_1) \land
(c_1 \lor c_2 \lor c_3)\land (b_1 \lor c_1 \lor c_3)$}
\label{figure:colorexamplesmall}
\end{figure}
\end{example}
In the following, we want to formalize how to build a color structure. We do so by formalizing how to add variables to $G$ without violating a property of the color structure. If two variables of a clause are contained in $G$ and we add the third variable to $G$, no variable of the clause is an edge label and the color structure yields an assignment where there is a clause where all variables are assigned to \texttt{false}. We say that a variable is \emph{locked} if it appears in a positive clause whose two remaining variables are both guards. A color is \emph{locked} if all variables of the color are locked. 

If only one variable of a clause is a guard and we want to choose a variable of the remaining two to become a guard, we choose the variable that is placed at the end point of the edge and hence contained in $N$ in order to comply the first property of the color structure. Otherwise, we need to swap the edge variable and the variable in $N$. Further, we can make sure that $G$ and $N \cup E$ are disjoint after we added a variable $x$ to $G$ if we add all variables from the clauses that contain $x$ to the sets $G,N$ and $E$. We abbreviate the process of swapping and adding variables by the term \emph{expansion}.

\begin{definition}
To \emph{expand a variable $l_s$}, proceed in the following steps:
\begin{itemize}
    \item While there is an edge $\{u_i,v_j\}^{l_s}$ in the color structure with edge label $l_s$ where $v_j$ is a guard, perform a swap between $u_i$ and $l_s$:
    \begin{itemize}
    \item $N:=N\cup \{l_s\}$
    \item If $\{u_i,v_j\}^{l_s}$ is the only edge in $E$ that contains $u_i$: $N:=N \setminus \{u_i\}$
    \item $E:=E \cup \{\{l_s,v_j\}^{u_i}\}$ and $E:=E \setminus \{\{u_i,v_j\}^{l_s}\}$
    \end{itemize}
\item For all clauses $c=(l_s \lor u_i \lor v_j)$ where $u_i$ is smaller than $v_j$, that are not contained in the color structure and that contain $l_s$ do: 
    \begin{itemize}
    \item Add an edge that is labeled with $u_i$: $E:=E \cup \{\{l_s,v_j\}^{u_i}\}$
    \item $N:=N \cup \{v_j\}$
    \end{itemize}
\item Then, $N:=N \setminus \{l_s\}$ and $G := G \cup \{l_s\}$
\end{itemize}
\end{definition}

Expansion of a variable can be seen as an update of the color structure that increases the number of guards. Conversely, we say that we \emph{unexpand} a variable if we remove the respective variable from the set $G$ and add it to $N$. We refrain from unexpanding variables that have just been expanded or the other way around. Note that the process of expanding a variable modifies the current color structure in a unique way.

Further, we define a measure of how close a color is to being locked. We call the \emph{lock number} of a variable of a color structure the number of relatives that have at least one locked occurrence or that are contained in a clause where one variable is a guard. By construction and the variable occurrence bound of four, the lock number ranges between zero and three. In Example \ref{example:colorsmall}, the lock number of variable $b_1$ is zero, while the lock number of $b_2$ and $b_3$ is one, since $b_1$ is locked between $a_1$ and $c_1$ and on top $b_1$ labels the edge $\{c_1,c_3\}^{b_3}$.  Intuitively, a high lock number indicates that the color of a variable is close to being locked. We therefore prioritize expanding variables with a high lock number. Similarly, we define the \emph{expand number} of a variable of a color structure as the number of locked colors that are gained if we expand the variable. 

\section{A Color Structure Algorithm}
Our first algorithmic approach for constructing a color structure is to greedily expand variables color by color.
The lock number serves as a  heuristic to avoid locking a color: if four variables of a color are locked, no satisfying assignment can be derived. If three variables of a color are locked, the only way to obtain a satisfying assignment is to expand the one remaining unlocked variable. The lock number ensures that this variable is prioritized for expansion.  Nevertheless, this greedy approach may still produce a locked color, and thus an unsatisfying assignment. In that case, we need to modify the current color structure to eliminate locked colors. If this fails, no satisfying assignment can be derived from the color structure. In particular, a color structure is \emph{satisfying} if each color has exactly one guard. The color structure depicted in Figure~2 (belonging to Example~\ref{example:8} below) is not satisfying, since no variable of color $g$ is a guard. The following lemma formalizes the correspondence between satisfying assignments and satisfying color structures. 
\begin{lemma}\label{lemma:satifiable-formula}
 A monotone 3-\textsc{Sat} formula $F$ in which  each variable occurs exactly once negated is satisfiable if and only if it admits a satisfying color structure.
 \end{lemma}  
\begin{proof}
Given a satisfying color structure $(G \cup N, E)$ for a formula~$F$, every variable is contained in $G \cup N \cup E$ and hence, carries an assigned truth value. Every positive clause of~$F$ is satisfied because at least one variable per clause belongs to~$E$, and all variables in~$E$ are set to \texttt{true}. All negative clauses of~$F$ are satisfied because every color contains a guard, which is set to \texttt{false}. 

Conversely, suppose $F$ is satisfiable, and let $\beta$ be a satisfying assignment. We construct a color structure by designating the false variables under~$\beta$ as guards and distributing the remaining variable occurrences into~$N$ and~$E$ according to the clause structure.  By construction, all properties of a satisfying color structure are fulfilled.
\end{proof}
In the following, we present the algorithm (Listing~\ref{lst:color-structure}) that takes a colored variant of a monotone 3-\textsc{Sat} formula $F$ where each variable occurs once negated and up to four times unnegated together as input and returns a color structure. The following example illustrates the individual steps of the algorithm. We resolve ties in line 3 of the algorithm by taking the smallest variable and ties in line 5 by taking the variable that occurs the least often and then, the highest variable in order to aim for a connected color structure. Other tie breaks are also possible.

\begin{lstlisting}[
caption={Color Structure Construction Algorithm},
label={lst:color-structure},
captionpos=t,
mathescape=true,
abovecaptionskip=-\medskipamount,
numbers=left
]
Set $Q = \{a_1\}$
WHILE $Q$ is not empty DO
    Expand unguarded variable in $Q$ with highest lock number
    WHILE there exists a locked color $u$ DO
        Choose variable $u_i$ with the lowest expand number 
        FOR each occurrence of $u_i$ DO
            Let $c=(u_i \lor v_j \lor l_s)$ be the clause containing $u_i$
            Among $v_j$ and $l_s$, unexpand the one with lower lock number
            Expand $u_i$
        END FOR
    END WHILE
    Add all variables in the color structure that are not locked to $Q$ 
    IF $Q$ is empty DO
        Pick a new unguarded variable $q_i$ that is not locked
        $Q = \{q_i\}$
    END IF  
    Remove variables $G$ and all their relatives from $Q$ 
END WHILE
RETURN Color structure constructed from the set of expanded variables
\end{lstlisting}\label{algo:ALGO}

\begin{example}\label{example:8}
Consider the formula  
\begin{align*}
&(a_1 \lor b_1 \lor f_1)
\land (a_1 \lor b_1 \lor f_2)
\land (a_1 \lor b_3 \lor c_1)
\land (a_2 \lor b_1 \lor c_3)
\land (a_3 \lor d_3 \lor g_3)\\
&\land (b_2 \lor c_1 \lor d_2)
\land (b_2 \lor f_2 \lor g_1)
\land (c_1 \lor d_1 \lor e_1)
\land (d_1 \lor e_2 \lor g_1)
\land (d_1 \lor e_2 \lor g_2)\\
&\land (c_2 \lor e_3 \lor g_3)
\land (e_2 \lor f_3 \lor g_3).
\end{align*}
First, we expand $a_1$ since the lock number of each variable is zero. Then, we have the following queue annotated with the current lock number in the second coordinate: $(b_1,1),(b_3,1),(c_1,0), (f_1,1)$ and $(f_2,1)$. Subsequently, we expand $b_1$ since it has the highest lock number and is the smallest variable. In the course of the algorithm, we expand the variables $a_1,\; b_1,\; c_1,\; d_1,\; e_2,\; f_3$ in this order. 
\begin{figure}[htbp]
    \centering
\begin{subfigure}{0.49\textwidth}
\centering
\resizebox{0.7\linewidth}{!}{
\begin{tikzpicture}[scale=0.8,  shorten >=0pt,node distance=3cm,on grid,auto] 
\node[state] (q2) {$c_1$};
\node[state] (q7) [left=of q2] {$a_1$};
\node[state] (q8) [left=of q7] {$b_1$};
\node (q3) [below left=of q8] {$a_2$};
\node (q6) [right=of q2] {$b_2$};
\node[state] (q9) [below=of q2] {$d_1$};
\node[state] (q10) [below=of q9] {$e_2$};
\node[state] (q11) [right=of q10] {$f_3$};

\path
(q8) edge node {$c_3$} (q3)
(q2) edge node {$b_3$} (q7)
(q7) edge[bend left] node {$f_1$} (q8)
(q9) edge node {$e_1$} (q2)
(q6) edge node {$d_2$} (q2)
(q10) edge[bend right] node {$g_1$} (q9)
(q10) edge[bend left] node {$g_2$} (q9)
(q10) edge node {$g_3$} (q11)
(q7) edge[bend right] node {$f_2$} (q8);
\end{tikzpicture}
}
\caption{Translating the colored formula from Example~\ref{example:8} to a color structure}
\label{figure:lockedstructure}
\end{subfigure}
    \hfill
\begin{subfigure}{0.49\textwidth}
\centering
\resizebox{0.9\linewidth}{!}{%
\begin{tikzpicture}[scale=0.8,  shorten >=0pt,node distance=3cm,on grid,auto] 

\node[state] (q2) {$c_1$};
\node[state] (q7) [left=of q2] {$a_1$};
\node[state] (q8) [left=of q7] {$b_1$};
\node (q3) [below left=of q8] {$a_2$};
\node (q6) [right=of q2] {$b_2$};
\node[state] (q9) [below=of q2] {$d_1$};
\node (q10) [below=of q9] {$e_2$};
\node[state] (q111) [right=of q9] {$f_3$};
\node[state] (q11) [below=of q111] {$g_3$};
\node (q4) [below left=of q11] {$a_3$};
\node[state] (q5) [below right=of q11] {$e_3$};

\path
(q3) edge node {$c_3$} (q8)
(q2) edge node {$b_3$} (q7)
(q9) edge node {$e_1$} (q2)
(q10) edge[bend right] node {$g_1$} (q9)
(q10) edge[bend left] node {$g_2$} (q9)
(q6) edge node {$d_2$} (q2)
(q11) edge node {$e_2$} (q111)
(q4) edge node {$d_3$} (q11)
(q5) edge node {$c_2$} (q11)
(q7) edge[bend left] node {$f_1$} (q8)
(q7) edge[bend right] node {$f_2$} (q8);
\end{tikzpicture}
}
\caption{Modified color structure for the formula from Example~\ref{example:8}}
\label{figure:finalstructure}
\end{subfigure}
\caption{Example Color Structure Construction Algorithm}
\end{figure}
This results in the color structure depicted in Figure~\ref{figure:lockedstructure}, locking the color $g$ and hence, the variables $g_1,\, g_2$, and $g_3$. Since all variables $g_1,\, g_2$, and $g_3$ occur exactly once, we expand $g_3$. The variable $g_3$ is locked between $e_2$ and $f_3$. Since the lock number of $e_2$ is one and the lock number of $f_3$ is two, we unexpand $e_2$ and then expand $g_3$. The outer while loop then terminates since we expanded the variable $e_3$ and with it the last color. This results in the color structure depicted in Figure~\ref{figure:finalstructure}. Now, we can extract the satisfying assignment: We set the guard variables 
$a_1,\; b_1,\; c_1,\; d_1,\; e_3,\; f_3,\; g_3$
to \texttt{false} and all remaining ones to \texttt{true}. Since there are exactly as many negative clauses as guards, we have that $Q$ is empty and the formula is satisfiable by Lemma~\ref{lemma:satifiable-formula}.
\end{example}

The outer while loop greedily expands a variable for each color. Since the inner while loop may create new locked colors, we record its operations in a \emph{state diagram}. An operation is depicted as rectangle and connected with the operation that is subsequent in the inner while loop. The inner while loop contains the operations expand or unexpand. The minimal size consists of six operations since we have that unexpansions are always followed by expansions and by definition, we do not immediately unexpand a variable that was just expanded or the other way around. The number of states in the state diagram depends on the input formula. In order to facilitate a later counting argument, we section an arbitrary state diagram into the base cycle and the $\ell$-block. The base cycle consists of the minimal size state diagram. We rename the variables such that the minimal size of a state diagram consists of the operations `expand $a_1$', `unexpand $b_1$', `expand $c_1$', `unexpand $a_1$', `expand $d_1$', `unexpand~$e_1$'. The remaining part of the state diagram consisting of the remaining $\ell$ expand operations and $\ell$ unexpand operations is called $\ell$-block. We rename the variables of the $\ell$-block that are expanded as $y_1^1, \dots, y_1^\ell$ and the variables $x_1^1, \dots, x_1^\ell$ that are unexpanded. Hence, the state diagram is labeled with variables of the colors $a,b,c,d,e,y^1, \dots, y^\ell, x^1, \dots, x^\ell$. This naming convention helps us to overview the states of the inner while loop (see Figure~\ref{figure:statediam}).

\begin{figure}[h]
\begin{center}
\resizebox{0.6\linewidth}{!}{%
\begin{tikzpicture}[
    state/.style={
        rectangle,
        draw,
        minimum size=1.8cm,
        align=center
    },
    arrow/.style={
        -{Stealth},
        thick
    },
    node distance=1.8cm
]
\node[state] (a) {expand\\$a_1$};
\node[state, right=of a] (b) {unexpand\\$b_1$};
\node[state, right=of b] (c) {expand\\$c_1$};
\node[below right=of c,
draw,
minimum size=1cm,
rectangle,
    label=right:{$\ell$-block}
] (repeat) {\dots};

\node[state, below=of c] (d) {unexpand\\$a_1$};
\node[state, left=of d] (e) {expand\\$d_1$};
\node[state, left=of e] (f) {unexpand\\$e_1$};

\draw[arrow] (a) -- (b);
\draw[arrow] (b) -- (c);
\draw[arrow] (c) -- (repeat);
\draw[arrow] (repeat) -- (d);
\draw[arrow] (d) -- (e);
\draw[arrow] (e) -- (f);
\draw[arrow, dotted] (f) -- (a);
\end{tikzpicture}}
\end{center}
\caption{State diagram of the inner while loop}
\label{figure:statediam}
\end{figure}
In order to prove correctness, we want to abstract from a concrete color structure given an input formula to the algorithm. Instead, we look at a hypothetical color structure that is constructed at a certain time in the algorithm. We specify this via the states of the state diagram. The state diagram fixes parts of the color structure due to the naming which variable is expanded or unexpanded when. We know that the variables that were expanded are guards and the variables that were unexpanded are edge labels connected to the guards. In particular, if `expand $a_1$' and `unexpand $b_1$' are consecutive, some clause must contain both $a_1$ and $b_1$. We call clauses \emph{connectors} that contain two variables from consecutive operations `expand' and `unexpand' because they connect the states in the state diagram. For each subsequent operation pair, we choose one specific connector if more than one clause contains both variables. In addition to connectors, there are clauses where we cannot specify its guards due to the naming of the state diagram but that contain variables that need to be locked somewhere in the current color structure. Throughout the algorithm, we only unexpand a variable if the corresponding color is locked in the first place. A color is locked if all (at most) four variables of the color have at least one occurrence that is locked. If we unexpand $b_1$ in order to expand $c_1$, we can assume two relatives $c_2$ and $c_3$ to be locked in the first place. For each unexpansion followed by an expansion of a variable, we call the relative variables of the variable \emph{detainee relatives}.

\begin{lemma}\label{lemma:countneedededgelab}
There are at least $6 +2\ell$ detainee relatives for the color structure $( G \cup N,E )$ constructed in the last state of the state diagram consisting of a base cycle and an $\ell$-block.
\end{lemma}
\begin{proof}
In the base cycle, we have that the colors $c,d$ and $a$ were locked. Consequently, six variables---$c_2,c_3,d_2,d_3,a_2$, and $a_3$---must be locked in addition to the three variables $a_1,c_1$ and $d_1$ that were locked in the course of the algorithm. In each iteration of the $\ell$-block, two additional variables $y_2^{i}$ and $y_3^{i}$ must be locked alongside~$y_1^{i}$. In total, this yields $6 + 2\ell$ variables that need to be locked for a cycle consisting of a base cycle and an $\ell$-block.
\end{proof}
So far, we have reasoned about the parts of the color structure at a certain state of the state diagram with fixed naming. In contrast, we can look at the parts of the color structure where we have not yet named the variables or fixed the location of the clauses. To illustrate the hypothetical color structure, imagine a color structure that consists of placeholders for the guards, edge labels and node labels were some are filled with variables and some are blank. The blank spots arise because no variable names have been assigned/specified in the inner while loop yet. We call edges whose labels have not yet been specified in the inner while loop \emph{prisoner spots}. A variable is locked if it is contained in a clause with two other guards and hence, prisoner spots are the locations where variables can be locked if the edge label is contained in a clause with two guards.\begin{lemma}\label{lemma:countfreeedgelabels}
If each variable occurs positively at most $k$ times, the color structure $( G \cup N,E )$ constructed in the last state of the state diagram consisting of a base cycle and an $\ell$-block has at most
$\lfloor \frac{3(k - 1) + (k - 1)\ell}{2} \rfloor$ prisoner spots.
\end{lemma}
\begin{proof}
First, we count all possible edge labels in the color structure $( G \cup N,E )$. We have three guards $a_1, c_1, d_1$, and $\ell$ additional ones due to the $\ell$-block, all of which are necessarily contained in $( G \cup N,E )$. Since each guard may occur at most $k$ times in the formula, we obtain $k(3+\ell)$ tuples of the form (guard, edge label). Here, we double count each edge label since two tuples (guard 1, edge label) and (edge label, guard 2) together correspond to a clause (guard 1, edge label, guard 2). Subsequently, there are at most $\lfloor \frac{k(3+\ell)}{2} \rfloor$ many edge labels in $( G \cup N,E )$. There might be less edge labels since there could be clauses that contain only one guard from $( G \cup N,E )$ and the remaining variables of the clause are not a guard or a guard that is not contained in $( G \cup N,E )$. Note that possibly, there must be clauses with variables that are not part of the loop color structure if $k$ is uneven and $\ell$ even. 

Second, we subtract from the number of all possible edge labels the edge labels that were already labeled in the course of the algorithm. Specifically, for each expansion in the cycle that is followed by an unexpansion, there must exist a connector clause that enforced the expansion in the first place. For example in the base cycle, we would only unexpand $b_1$ and expand $c_1$ due to a clause $(a_1 \lor c_1 \lor b_1)$ which results in at least one edge labeled with $c_1$. Since a clause contains three variables, for each operation pair, we  subtract half a clause from the total number of prisoner spots. We therefore subtract $\frac{3}{2}$ for the edges originating from the clauses $(a_1 \lor c_1 \lor b_1)$, $(y_1^i \lor d_1 \lor a_1)$ and $(d_1,a_1,e_1)$, and subtract a further~$\frac{\ell}{2}$ since each expansion in the $\ell$-block must be connected to a subsequent expansion in the $\ell$-block. In sum, there are at most
$\lfloor \frac{k(3+\ell) - 3 - \ell}{2} \rfloor = \lfloor \frac{3(k - 1) + (k - 1)\ell}{2} \rfloor$ prisoner spots.
\end{proof}
In the following, we use the counting arguments to prove correctness of the Color Structure Construction Algorithm.
\begin{theorem}\label{theorem:main}
Every monotone 3-\textsc{Sat} formula~$F$ in which each variable occurs once negated and at most four times unnegated is satisfiable. A satisfying assignment can be constructed in time $\mathcal{O}(n \cdot m)$, where $n$ is the number of negative clauses and $m$ is the number of positive clauses of~$F$.
\end{theorem}
\begin{proof}
Without loss of generality, we assume that each variable in a negative clause occurs in at least one positive clause. If this is not the case, we can immediately  satisfy the negative clause by assigning  \texttt{false} to any variable that does not appear in a positive clause and \texttt{true} to all remaining variables. In addition, we recall that it is only beneficial to set at most one variable per negative clause to \texttt{false}. As a result, we assume that no two variables of a positive clause occur in the same negative clause since then, the positive clause is automatically satisfied. 

First, we show that the algorithm terminates. The outer while loop terminates once a variable has been marked as a guard for each color. In each iteration, one differently colored variable is added to $G$ since in line 13--16 the algorithm adds variables to $Q$ as long as there are colors that do not have a guard yet. The inner while loop may exchange variables in~$G$, but does not decrease the number of guards. After all colors have been expanded, $Q$ is empty since all variables are in $G$ or are relatives of variables in $G$. For termination of the inner while loop, we look at the state diagram. The maximum size of the state diagram is reached when all clauses of the formula are in the color structure that is modified in the inner while loop. Afterwards, we modify the containment of the variables that was modified before. We depict this with a dotted line in the state diagram (Figure \ref{figure:statediam}) and show that we exit the while loop before entering a state that was entered before.

Let $( G \cup N,E )$ be the current color structure in the last state of the state diagram restricted to the part of the color structure that contains the variables $$\mathcal{C} = \{a_1,c_1,d_1,y_1^1, y_1^2, \dots ,y_1^\ell\}$$ or is locked by variables in $\mathcal{C}$. In the following, we compare the maximum number of prisoner spots with the minimum number of detainee relatives. If not all detainee relatives fit into the prisoner spots, for each color there are variables that are not locked and hence, can be expanded. Subsequently, the algorithm stops. Due to Lemma \ref{lemma:countneedededgelab}, there are $6 + 2\ell$ detainee relatives. Furthermore, the detainee relatives must be locked between the variables in $\mathcal{C}$ if we have a state diagram consisting of a base cycle and an $\ell$-block: Suppose there exists a color $h \in \mathcal{C}$ with a locked variable where one guard is not contained in $\mathcal{C}$. Then, this variable has an expand number of zero and hence, is prioritized in the algorithm, and the inner while loop is exited after this iteration. But then, the color structure is not the one constructed in the last state of the state diagram. Due to Lemma \ref{lemma:countfreeedgelabels}, we have at most $\lfloor 4.5 + 1.5\ell \rfloor$ prisoner spots for variables that occur positively at most four times, and $6+2\ell$ prisoner spots for variables that occur positively at most five times. Thus, a cycle in the state diagram is possible when each variable occurs positively at most five times, but impossible when occurrences are bounded by four.

Further, we want to elaborate on why the algorithm is well-defined. First, note that we can always expand a color (line 3) without violating property one of the definition of a color structure since variables of positive clauses have different colors. We have property two since in line 17, we remove variables with colors that already have a guard and property three is given due to the expansion procedure. Similarly, the current color structure is a valid color structure before and after each iteration of the inner while loop because unexpanding all occurrences of a variable unlocks a variable and hence, allows the expansion of it yielding a valid color structure. We have that the inner while loop terminates in $\mathcal{O}(m)$ after---in the worst case---all clauses that have been added as edges in the color structure were modified. The expand number can be computed in constant time since the variable occurs only constantly many clauses. The outer while loop terminates in $\mathcal{O}(n)$ once there is a guard for each color. By Lemma \ref{lemma:satifiable-formula}, this yields an $\mathcal{O}(n \cdot m)$ algorithm producing a satisfying assignment for any monotone 3-\textsc{Sat} formula with at most four positive occurrences and one negative occurrence per variable.
\end{proof}
As a result, we obtain an algorithm that iteratively builds a color structure from a given colored formula and thereby derives a satisfying assignment. Together with the equivalence proved by Darmann and D\"ocker \cite[Theorem 1]{janoschPaper} and the aforementioned hardness result showing that \textsc{Monotone 3-Sat}-$(k, 1)$ is NP-complete for all $k \geq 5$ {\cite[Corollary 7]{janoschPaper}, we obtain the following dichotomy:
\begin{corollary}
The problems \textsc{Monotone 3-Sat}-($\leq k$,1) and \textsc{Monotone 3-Sat}-($k$,1) are NP-hard for $k \geq 5$ and trivial for $1 \leq k \leq 4$. 
\end{corollary}
\section{Conclusion}

In sum, we introduced the concept of a color structure for encoding assignments for formulas of the type \textsc{Monotone 3-Sat}-$(\leq k,1)$ with $k\geq 1$. We presented a polynomial-time algorithm that translates a formula of this type into a color structure and used it to establish the existence and construction of a satisfying assignment for $k\leq 4$. The correctness proof of the algorithm 
relies on a counting argument: For $k \leq 4$, as shown in the proof of Theorem \ref{theorem:main}, each color has variables that are not locked and can therefore be expanded, which guarantees that the algorithm terminates with a satisfying assignment.
For $k\geq 5$, this argument no longer applies, since there are enough prisoner spots for all detainee relatives, so that the loop may not be left. This breakdown is not incidental: \textsc{Monotone 3-Sat}-$(\leq k,1)$ is NP-hard for~$k\geq 5$, which in particular implies the existence of unsatisfiable instances. If the counting argument still held, it would show that every instance is satisfiable, contradicting this hardness result. In this sense, the color structure allows us to pinpoint exactly where the argument breaks down as~$k$ increases from 4 to 5, matching the transition in complexity. This suggests that color structures capture relevant structural information about such formulas beyond the specific algorithm presented here, and we believe the concept may prove fruitful for other \textsc{Sat}-solving approaches as well.

\bibliography{SATproject}
\end{document}